\documentclass[12pt]{article} 
\usepackage{amsmath}
\usepackage{amssymb}
\usepackage{epsfig}

\title{\textbf{Quantum Whispers}}
\author{\textit{Lucien Hardy}\\
Clarendon Laboratory, University of Oxford\thanks{
Clarendon Laboratory, 
Department of Physics, 
University of Oxford,
OX1~3PU Oxford,
United Kingdom. Email address: \texttt{hardy@mildred.physics.ox.ac.uk}}\\
~ \\
\textit{Wim van Dam}\\
Centre for Quantum Computation, University of Oxford\thanks{
Centre for Quantum Computing,
Clarendon Laboratory, 
Department of Physics, 
University of Oxford,
OX1~3PU Oxford,
United Kingdom.
Quantum Computing and Advanced Systems Research,
C.W.I., 
P.O.~Box~94079, 
NL--1090~GB Amsterdam, The Netherlands. Email address: 
\texttt{wimvdam@mildred.physics.ox.ac.uk}}\\
Quantum Computing and Advanced Systems Research, CWI}

\begin{document}

\maketitle

\begin{abstract}
It is shown that with the use of entanglement a specific 
two party communication task can be done with a systematically smaller 
expected error than any possible classical protocol could do.  
The example utilises the very tight correlation between separate 
spin measurements on a singlet state for small differences in the angles 
of these two measurements. 
An extension of this example to  many parties arranged in a row with only 
local, one--to--one communication (whispering) is then considered.
It is argued that in this scenario there exists no reliable classical
protocol, whereas in the quantum case there does.
\end{abstract}
\newpage
\section{Introduction}
Quantum entanglement is used in various ways, both to illustrate 
fundamental aspects of quantum mechanics and in information processing 
applications.  In the first category are the Einstein 
Podolsky Rosen argument and Bell's theorem.  In the latter category are 
teleportation, quantum cryptography, and quantum computing.  Recently, 
another such application has been found by Cleve and Buhrman\cite{Cleve1}, 
namely that of `entanglement enhanced classical communication' or 
`quantum communication' for short.
It has been shown by various authors\cite{Buhrman1,Buhrman2,Cleve1,vDam}
that the \emph{communication complexity\/} of certain
communicational tasks can be reduced with the use of entanglement.
This has to be understood as follows.

Several spatially separated parties $A,B,C,\dots$ are each given 
their personal data $x_1,x_2,x_3,\dots$, which is initially unknown to the
other parties. Next, the parties communicate among each other
by means of classical bits with the objective that one party (say $A$) 
can announce the value of a previously given function $f$ on the input 
values $x_1,x_2,\ldots$ after a minimal amount of communication.
That is, $A$ wants to know the specific 
function value $f(x_1,x_2,\ldots)$ after the protocol with the least 
possible number of bits communicated during the protocol.
The following stages are to be distinguished in the whole procedure:  
\begin{description}
\item[stage 1:]{Before being given the data $x_i$ the parties are allowed to 
communicate as much as they like (for example to share random numbers).}  
\item[stage 2:]{After they have been given the data $x_i$, the parties must 
communicate as little as possible.}
\end{description}
The \emph{communication complexity\/} of a function $f$ is 
the minimum amount of communication (measured in bits) in 
stage 2 that is necessary to determine the value of $f(x_1,x_2,\ldots)$.  
The \emph{quantum communication complexity\/} is defined in the same way
as the classical communication complexity except that now, during stage 1, 
the parties are also allowed to share quantum particles (qubits) that 
are entangled. 

By considering an example involving three parties, Cleve and 
Buhrman\cite{Cleve1} showed that for a specific function $f$, less 
communication is required in the quantum case than in the classical case. 
This is surprising because quantum entanglement cannot be used to send 
a signal from one place to another, nor can it even compress information
when using a classical communication channel\cite{Werner}.
However, when embedded into an appropriate task which involves communication, 
the nonlocal properties of an entangled state can allow a reduction in the 
amount of communication required. 
Unlike the other applications of quantum mechanics mentioned above, this 
application really does make explicit use of the Bell type nonlocality 
of entangled states.  Indeed, one can say that whenever we have an example 
of a function where a quantum protocol beats the best possible classical 
protocol (in terms of communication complexity), 
it can be re-interpreted as a proof of the nonlocality of quantum 
mechanics (the converse statement does not appear to be true though). 
 
After the initial article by Cleve and Buhrman, several other results 
on quantum communication were obtained. Some of them expanded the
difference in communication complexity between the quantum and the
classical case \cite{Buhrman1,Buhrman2,vDam}, whereas others showed that 
for certain functions there can be no essential difference between the 
two scenario's \cite{Buhrman2,Cleve2}. See also the notion of 
telecomputation\cite{Ekert,Grover} for related work.

In this paper we will take a slightly different approach to the 
problem.  Rather than calculating the amount of communication required to 
compute the function value, we will allow only a certain amount 
of communication and calculate the expected error rate under this restriction. 
(A similar approach to probabilistic communication is taken 
in \cite{Buhrman1}, Section 3.)
At first we will analyse a task for two parties $A$ and $B$. 
It will be shown that if $A$ and $B$ are allowed to share a singlet state
the expected error-rate
can be systematically lower than the classical case when they are not
allowed to share such an entangled state.
The essence of this result relies heavily on the correlation between 
spin measurements that differ either by small angles, or by approximately
$\pi$.
This typical behaviour has been considered by various 
authors\cite{Braunstein,Hardy,Squires},
and can be used to run a proof of Bell's theorem by invoking an effect 
mathematically analogous to the \emph{quantum Zeno effect.}

We will also consider an extension of this situation in which a large 
number of parties, $A,B,\dots$ are 
arranged in a row and communicate pairwise (i.e. $A$ and $B$, $B$ and $C$, 
etc.) It is argued that in this scenario the quantum case allows for
a correct protocol (with high probability), whereas in the classical
case no reliable protocol exists to calculate the function value.
The similarity of this situation to the well known game of 
Chinese whispers gives this paper its title.

\section{A Two Party Communication Problem}
\begin{figure}
\begin{picture}(0,0)%
\includegraphics{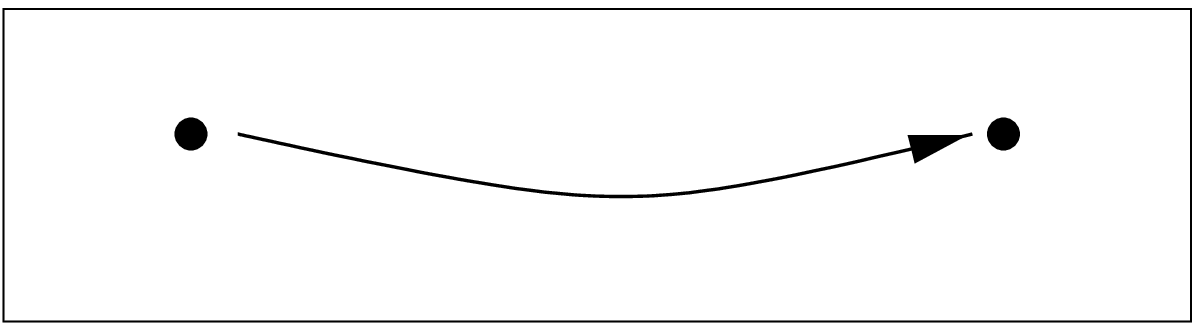}%
\end{picture}%
\setlength{\unitlength}{3947sp}%
\begingroup\makeatletter\ifx\SetFigFont\undefined
\def\x#1#2#3#4#5#6#7\relax{\def\x{#1#2#3#4#5#6}}%
\expandafter\x\fmtname xxxxxx\relax \def\y{splain}%
\ifx\x\y   
\gdef\SetFigFont#1#2#3{%
  \ifnum #1<17\tiny\else \ifnum #1<20\small\else
  \ifnum #1<24\normalsize\else \ifnum #1<29\large\else
  \ifnum #1<34\Large\else \ifnum #1<41\LARGE\else
     \huge\fi\fi\fi\fi\fi\fi
  \csname #3\endcsname}%
\else
\gdef\SetFigFont#1#2#3{\begingroup
  \count@#1\relax \ifnum 25<\count@\count@25\fi
  \def\x{\endgroup\@setsize\SetFigFont{#2pt}}%
  \expandafter\x
    \csname \romannumeral\the\count@ pt\expandafter\endcsname
    \csname @\romannumeral\the\count@ pt\endcsname
  \csname #3\endcsname}%
\fi
\fi\endgroup
\begin{picture}(5724,1524)(-11,-1273)
\put(1426,-961){\makebox(0,0)[lb]{\smash{\SetFigFont{12}{14.4}{rm}1 classical 
bit of communication}}}
\put(826,-136){\makebox(0,0)[lb]{\smash{\SetFigFont{12}{14.4}{rm}$A$}}}
\put(4696,-661){\makebox(0,0)[lb]{\smash{\SetFigFont{12}{14.4}{rm}$y$}}}
\put(4726,-136){\makebox(0,0)[lb]{\smash{\SetFigFont{12}{14.4}{rm}$B$}}}
\put(751,-661){\makebox(0,0)[lb]{\smash{\SetFigFont{12}{14.4}{rm}$x$}}}
\end{picture}
\caption{Schematic view of the two party protocol: Alice (who has 
value $x$) is allowed to communicate 1 classical bit to Bob 
(who has value $y$).
He then has to answer the ``jump or no jump?'' question.}
\label{schematic}
\end{figure}

Now let us introduce the communication problem for the two parties $A$ 
and $B$, or Alice and Bob.  During stage 1 they are allowed to 
exchange information and, in the quantum 
case, quantum entanglement.  After this stage they are each given a 
number $x$ and $y$ respectively, 
where $x,y \in \{0,1,2,\dots,2N-1\} $. We will employ
modular arithmetic so that these numbers can be thought of as being on a
circle with $2N$ dots around it. The person providing 
these numbers makes a promise.  He promises that either 
\begin{equation}\label{promise}
\begin{array}{lrcll} 
& x-y & \in &  \{-1,0,+1\} & \qquad 
\textrm{called ``no jump''} \\
\text{or ~ ~} & x-y & \in & \{N-1,N,N+1\} & \qquad 
\textrm{called ``jump'',} \\
\end{array}
\end{equation}
where arithmetic is understood to be modulo $2N$ as already stated. 
\begin{figure}
\begin{picture}(0,0)%
\includegraphics{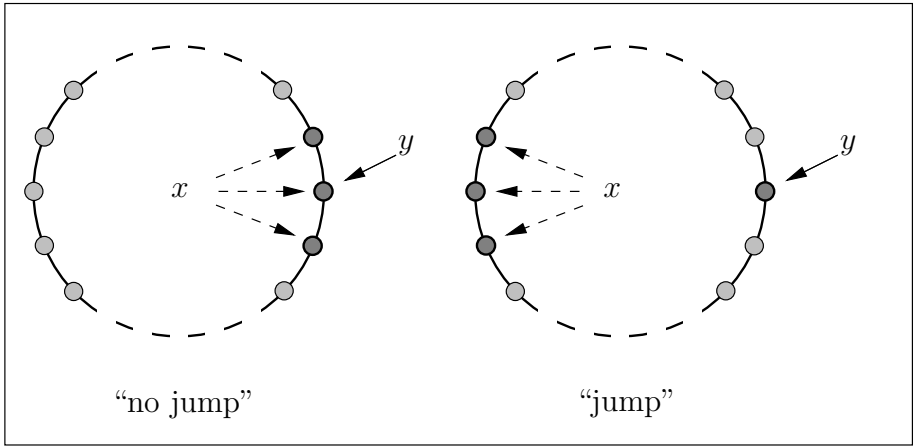}%
\end{picture}%
\setlength{\unitlength}{3947sp}%
\begingroup\makeatletter\ifx\SetFigFont\undefined
\def\x#1#2#3#4#5#6#7\relax{\def\x{#1#2#3#4#5#6}}%
\expandafter\x\fmtname xxxxxx\relax \def\y{splain}%
\ifx\x\y   
\gdef\SetFigFont#1#2#3{%
  \ifnum #1<17\tiny\else \ifnum #1<20\small\else
  \ifnum #1<24\normalsize\else \ifnum #1<29\large\else
  \ifnum #1<34\Large\else \ifnum #1<41\LARGE\else
     \huge\fi\fi\fi\fi\fi\fi
  \csname #3\endcsname}%
\else
\gdef\SetFigFont#1#2#3{\begingroup
  \count@#1\relax \ifnum 25<\count@\count@25\fi
  \def\x{\endgroup\@setsize\SetFigFont{#2pt}}%
  \expandafter\x
    \csname \romannumeral\the\count@ pt\expandafter\endcsname
    \csname @\romannumeral\the\count@ pt\endcsname
  \csname #3\endcsname}%
\fi
\fi\endgroup
\begin{picture}(5724,2799)(-11,-2988)
\put(1051,-1411){\makebox(0,0)[lb]{\smash{\SetFigFont{12}{14.4}{rm}$x$}}}
\put(5251,-1111){\makebox(0,0)[lb]{\smash{\SetFigFont{12}{14.4}{rm}$y$}}}
\put(2476,-1111){\makebox(0,0)[lb]{\smash{\SetFigFont{12}{14.4}{rm}$y$}}}
\put(3766,-1413){\makebox(0,0)[lb]{\smash{\SetFigFont{12}{14.4}{rm}$x$}}}
\put(676,-2761){\makebox(0,0)[lb]{\smash{\SetFigFont{12}{14.4}{rm}``no jump''}}}
\put(3593,-2758){\makebox(0,0)[lb]{\smash{\SetFigFont{12}{14.4}{rm}``jump''}}}
\end{picture}
\caption{Explanation of the communication problem: 
``no jump'' means that $x$ equals $y$ or is adjacent to it,
and ``jump'' stands for the situation where $x$ is either opposite $y$
or adjacent to that opposite position.}
\label{jumpnojump}
\end{figure}
In the case of ``no jump'' $x$ is either equal or immediately adjacent to 
$y$.  In the case of ``jump'' $x$ is either opposite $y$ on the 
circle or either side of that opposite position (see Figure \ref{jumpnojump}). 
During stage 2 party $A$ is allowed to send \emph{one classical bit\/} 
of information (say either $+1$ or $-1$) to party $B$ after which Bob
has to announce whether there has been no jump or a jump. 
This situation is depicted in Figure \ref{schematic}.
(We could write this task down as being equivalent to evaluating an 
appropriately defined function on a restricted domain.)
We assume that $N\geq3$ such that the cases of jump and no jump are 
mutually exclusive.
As is customary in communication complexity theory, we define the 
`expected error-rate' as the the expected error for the worst possible
probability distribution over allowed input numbers $x$ and $y$
(worst case scenario assumption). An error occurs when
$B$ states incorrectly whether there has been a jump or not.
We will see that 
\begin{enumerate} 
\item{For a described quantum protocol, the expected error-rate is 
proportional to $1/N^2$.}
\item{In the classical case, the expected error-rate will always
be equal to or greater than $k/N$ where $k$ is a constant.} 
\end{enumerate}
This means that for sufficiently large $N$, the quantum protocol 
outperforms all possible classical protocols. 
 
\subsection{The Quantum Protocol}
The quantum case is easier to prove so we will start there. 
During stage 1  of the protocol the two parties share a singlet state:
\begin{eqnarray}
|\psi\rangle & = &
\frac{1}{\sqrt{2}}(
{|+\rangle}_{\mathrm{A}}{|-\rangle}_{\mathrm{B}} -
{|-\rangle}_{\mathrm{A}}{|+\rangle}_{\mathrm{B}}).
\end{eqnarray}
Then, at the start of the second stage, the two parties receive their 
respective numbers $x$ and $y$.  According to those numbers, 
they define $\theta={2\pi x/2N}$ and $\phi={2\pi y/2N}$.  
Now Alice measures the spin on her particle 
along the angle $\theta$ to the $z$-axis in the $xz$ plane.  At the same 
time Bob measures the spin on his particle along the angle $\phi$ to the 
$z$-axis in the $xz$ plane.  Each of them will get a result $\pm 1$ as 
the outcome of this measurement.  To simplify matters for further analysis,
Bob changes the sign of his outcome so that 
$\pm 1$ becomes $\mp 1$ (this means that if $\theta=\phi$ then 
Alice and Bob have the same outcome).  Given 
this we can easily show that for the singlet state we have: 
\begin{eqnarray} 
\text{Prob}(\text{same}|\theta,\phi) 
& = & 
\frac{1}{2}(1+\cos(\theta-\phi)) \\
\text{Prob}(\text{opposite}|\theta,\phi)
& = & 
\frac{1}{2}(1-\cos(\theta-\phi)),
\end{eqnarray} 
where Prob(same) is the probability that Alice and Bob's outcomes are the 
same and Prob(opposite) is the probability that their outcomes are 
opposite. 

In the case where there is no jump we have 
\begin{eqnarray} \label{angle1}
\left|{\theta-\phi}\right| 
& \leq &  \frac{\pi}{N}
\end{eqnarray}
and hence 
\begin{eqnarray}\label{samenj}
\text{Prob}(\text{same}|\text{no jump}) 
& \geq & 
1-{\left(\frac{\pi}{2N}\right)}^2 \\ \label{oppositenj}
\text{Prob}(\text{opposite}|\text{no jump}) 
& \leq & 
{\left(\frac{\pi}{2N}\right)}^2. 
\end{eqnarray}
In the case where there is a jump we have
\begin{equation} \label{angle2}
\pi-\frac{\pi}{N} ~ \leq~ 
\left|{\theta-\phi}\right| ~ \leq ~ 
\pi+\frac{\pi}{N}
\end{equation}
and hence
\begin{eqnarray}\label{oppositej}
\text{Prob}(\text{opposite}|\text{jump}) 
& \geq & 1-{\left({\frac{\pi}{2N}}\right)}^2 \\ \label{samej}
\text{Prob}(\text{same}|\text{jump}) 
& \leq & {\left(\frac{\pi}{2N}\right)}^2 
\end{eqnarray} 
Therefore Alice and Bob can adopt the following protocol.
Alice sends the result of her spin measurement ($\pm1$) to Bob along 
the classical channel. 
This constitutes the one allowed bit of classical communication. Next,
Bob multiplies this result by his own spin measurement outcome. 
If the result of this is +1 (i.e. Alice and Bob's outcomes are the 
same) then it follows from Equations
(\ref{samenj}), (\ref{oppositenj}), (\ref{oppositej}), and (\ref{samej})
that it is most likely that there was no jump.
Hence Bob announces that there was no jump. 
If the result of this multiplication is $-1$ 
(i.e. the two results are opposite), Bob announces that there
has been a jump.

For both cases the worst possible angle deviation is $\pi/N$ 
(Equations (\ref{angle1}) and (\ref{angle2})), and hence the expected
error less than or equal to $({\pi/2N})^2$. 
The above described quantum protocol has therefore an expected error-rate 
of the order $1/N^2$: 
\begin{eqnarray} \label{quantum_result}
{\rm Error_{quantum}} & \leq & {\left(\frac{\pi}{2N}\right)}^2.
\end{eqnarray} 
For large $N$, this probability tends very quickly to zero. 

\subsection{The Best Possible Classical Protocol}
In analysing the classical case we want to be sure that we have found 
the minimum possible error.  This means that we must be more careful and the 
analysis is correspondingly more detailed than in the quantum case. 

We will start by assuming that the provider of the numbers $x$ and 
$y$ distributes them evenly over the possibilities 
such that they satisfy the promise in Equation (\ref{promise}).  
The probability that $y$ takes some particular value in the set 
$\{0,\ldots,2N-1\}$ is therefore ${1/2N}$.  
For this particular value, Bob knows that there are 
six possible values for $x$.  Since the distribution is uniform, each of 
these possible values is equally probable with probability $1/6$. 
Alice's task is to communicate one bit (either $+1$ or $-1$) to Bob 
in such a way that Bob has a good chance of correctly saying whether 
there has been a jump or not.  We assume that Alice behaves in a 
deterministic way (though later we will also discuss the indeterministic 
case).  Thus, the best Alice can do is to look up the value of 
some function $g(x)=\pm 1$ and communicate this to Bob via the one bit 
classical channel (this corresponds to the deterministic case because 
a for a given $x$ Alice will always send the same message).  
Bob can also know the form of the function $g$ and hence when he 
receives a $+1$ he knows which subset Alice's $x$ is in, and similarly 
for the disjunct set if he receives a $-1$.  We will rephrase 
this into a colouring problem. 
\begin{figure}
\begin{picture}(0,0)%
\includegraphics{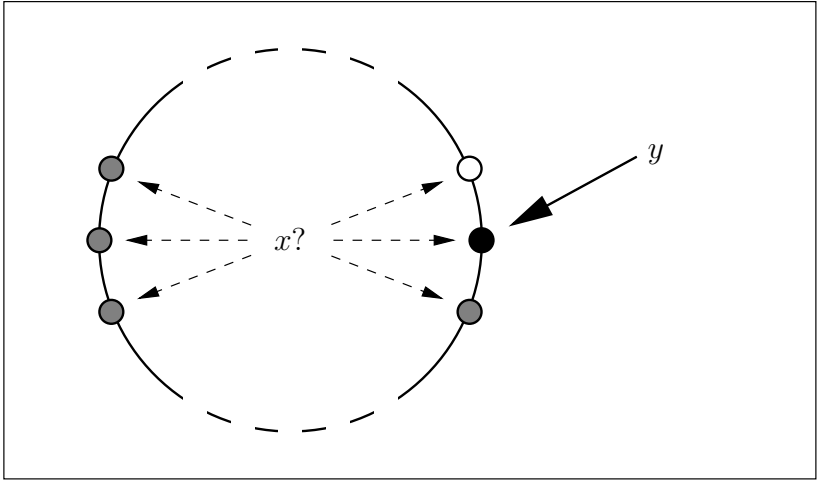}%
\end{picture}%
\setlength{\unitlength}{3947sp}%
\begingroup\makeatletter\ifx\SetFigFont\undefined
\def\x#1#2#3#4#5#6#7\relax{\def\x{#1#2#3#4#5#6}}%
\expandafter\x\fmtname xxxxxx\relax \def\y{splain}%
\ifx\x\y   
\gdef\SetFigFont#1#2#3{%
  \ifnum #1<17\tiny\else \ifnum #1<20\small\else
  \ifnum #1<24\normalsize\else \ifnum #1<29\large\else
  \ifnum #1<34\Large\else \ifnum #1<41\LARGE\else
     \huge\fi\fi\fi\fi\fi\fi
  \csname #3\endcsname}%
\else
\gdef\SetFigFont#1#2#3{\begingroup
  \count@#1\relax \ifnum 25<\count@\count@25\fi
  \def\x{\endgroup\@setsize\SetFigFont{#2pt}}%
  \expandafter\x
    \csname \romannumeral\the\count@ pt\expandafter\endcsname
    \csname @\romannumeral\the\count@ pt\endcsname
  \csname #3\endcsname}%
\fi
\fi\endgroup
\begin{picture}(5124,3024)(289,-3973)
\put(4351,-1936){\makebox(0,0)[lb]{\smash{\SetFigFont{12}{14.4}{rm}$y$}}}
\put(2003,-2514){\makebox(0,0)[lb]{\smash{\SetFigFont{12}{14.4}{rm}$x$?}}}
\end{picture}
\caption{Given the fact that Bob knows $y$, there are six 
equally likely possibilities for the Alice's value $x$. Here we have
chosen $y$ such that for some allowed values of $x$ Alice will send $+1$ 
(black) and for other possible values $-1$ (white).}
\label{blackwhiteadj}
\end{figure}  

\begin{figure}
\begin{picture}(0,0)%
\includegraphics{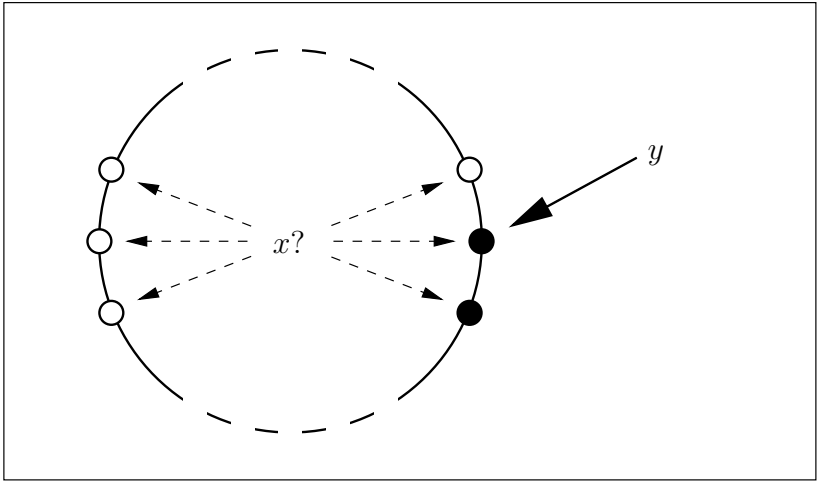}%
\end{picture}%
\setlength{\unitlength}{3947sp}%
\begingroup\makeatletter\ifx\SetFigFont\undefined
\def\x#1#2#3#4#5#6#7\relax{\def\x{#1#2#3#4#5#6}}%
\expandafter\x\fmtname xxxxxx\relax \def\y{splain}%
\ifx\x\y   
\gdef\SetFigFont#1#2#3{%
  \ifnum #1<17\tiny\else \ifnum #1<20\small\else
  \ifnum #1<24\normalsize\else \ifnum #1<29\large\else
  \ifnum #1<34\Large\else \ifnum #1<41\LARGE\else
     \huge\fi\fi\fi\fi\fi\fi
  \csname #3\endcsname}%
\else
\gdef\SetFigFont#1#2#3{\begingroup
  \count@#1\relax \ifnum 25<\count@\count@25\fi
  \def\x{\endgroup\@setsize\SetFigFont{#2pt}}%
  \expandafter\x
    \csname \romannumeral\the\count@ pt\expandafter\endcsname
    \csname @\romannumeral\the\count@ pt\endcsname
  \csname #3\endcsname}%
\fi
\fi\endgroup
\begin{picture}(5124,3024)(289,-3973)
\put(4351,-1936){\makebox(0,0)[lb]{\smash{\SetFigFont{12}{14.4}{rm}$y$}}}
\put(1996,-2528){\makebox(0,0)[lb]{\smash{\SetFigFont{12}{14.4}{rm}$x$?}}}
\end{picture}
\caption{A specific colouring of the four grey dots in Figure 
\ref{blackwhiteadj}. The analysis in the text shows that for this example,
Alice and Bob have an error probability of $1/6$ when trying to decide if 
$x$ and $y$ make a jump or not.}
\label{blackwhiteadj2}
\end{figure}

Each value of $x$ gives rise to a $+1$ or a $-1$ for $g(x)$ 
which we will colour as black and white 
respectively.  Thus, we can colour each dot $x$ on the circle representing 
the values of $g(x)$. 
 Except in the trivial case where all dots are either 
black or white, there must be at least one place on the 
circle where black and white dots are adjacent. 
 Consider these two dots, one that is adjacent to the two and
the three opposite dots making six dots in total. 
Assume also that $y$ is one of the two non-equally coloured pair,
as shown in Figure \ref{blackwhiteadj}. 
In this figure there are two dots of known colour and four of unstated 
colour.  There are therefore 16 ways of colouring these remaining dots. 
 We can consider each way individually and calculate the expected error 
for Alice and Bob in determining whether there has been a jump or not.  
One example of such a colouring is shown in Figure \ref{blackwhiteadj2}.
Imagine that the value of $y$ is as indicated by
the arrow in this picture (this occurs with probability $1/2N$). 
The probability that Alice sends a $+1$ (black) is $2/6$ (because two 
of the six dots are black) and the probability that she sends a $-1$ 
(white) is $4/6$.  
If Bob receives a $-1$ (i.e. white) from Alice then he cannot be sure 
whether $x$ corresponds to a jump or not, though it is more likely 
that it corresponds to a jump so he will announce that there has been 
a jump. 
The probability of error corresponds to the one white on the no jump 
side.  As one of the four whites are on the no jump side the probability 
of error is $1/4$.  
If Bob receives a $+1$ (i.e. black) from Alice then he can be sure that 
$x$ corresponds to no jump since both of the blacks are on the no jump 
side. Hence this probability of error is zero.  
Taking all this into account, the contribution of this colouring 
combination with this specific value of $y$ to the overall error is: 
\begin{eqnarray} \label{figure_error}}
{\mathrm{error}}_{\mathrm{Figure~\ref{blackwhiteadj2}}
&=&  
\frac{1}{2N}\left[{
\left({\frac{4}{6}}\right)\left({\frac{1}{4}}\right) +
 \left({\frac{2}{6}}\right) \left({\frac{0}{2}}\right)
}\right].
\end{eqnarray} 
Having performed one analysis of this kind, 
it is a simple matter to go through the remaining 15 colouring 
possibilities.  When doing this, one sees that the particular example 
given above is one of the ways of colouring that gives rise to the 
smallest error.

When colouring the $2N$ dots around the circle there are the following
two possible cases. If the dots are coloured uniformly black (or white)
then Bob does not receive any information from Alice, hence the
probability of error will be $1/2$. It is also clear that the related case 
of just one black (or white) dot will lead to a high error-rate.
In the case where there are at least two black and two white 
dots there will be at least $4$ instances of situation
similar to Figure \ref{blackwhiteadj} and the above analysis. 
Each with a least expected error as expressed in 
Equation (\ref{figure_error}).
Hence, we can conclude that the overall error-rate of any classical
protocol will be of the order $1/N$: 
\begin{eqnarray} \label{clas_result} 
\mathrm{Error}_{\mathrm{classical}} & \geq & \frac{1}{3 N}~.
\end{eqnarray} 
The reason for this classical error can be intuitively understood in 
the following way.  Alice's strategy will 
involve partitioning the circle into at least two parts where each part
is coloured either black or white. 
 As long as the values of $x$ and $y$ are well inside
these sections there will be no error.  However, if $x$ and $y$ are 
at the boundary between these two parts then there is likely to be an 
error.  The probability of being at such a boundary on the circle 
goes as $1/N$, which explains the above result. 

It is interesting to note that if Alice
is allowed to send one trit of communication (black, white, or blue)
then an appropriate colouring will lead to no errors.  For example the
circle can be coloured in the following way: For
$x\in\{2N-1,0,1,2\}$ put a blue dot; for $x\in\{3,\dots,N\}$ put a 
white dot, and for $x\in\{N+1,\dots,2N-2\}$ put
a black dot. A little thought shows that this arrangement will lead to
no errors.  It can be shown that one trit corresponds to an average of
approximately $1.5$ bits of communication by using the unambiguous 
set of code-words `$0$',`$10$', and `$11$'\cite{Cover}.
Thus, we can say that, in some sense,  the quantum state is substituting
for an average of not more than approximately half a bit of classical
information.

\subsection{Remarks on Randomisation}
There are two subtleties concerning randomisation which should be 
discussed before honestly accepting the results of Equations 
(\ref{quantum_result}) and (\ref{clas_result}).

First we have to ask whether it is possible that Alice and Bob 
could do better if Alice were to send $+1$ and $-1$ according to some
randomised protocol. The answer to this question is in short: 
this can not be the case because we 
fixed the distribution of $x$ and $y$ (the homogeneous distribution)
before we proved the error-rate. The reasoning behind this statement
goes as follows.
If Alice and Bob would employ some randomisation in their protocol,
then the description of the protocol \emph{in situ\/}
will depend on some random numbers $r$. 
For a fixed distribution on the input data, some values of $r$ 
(denoted by $r'$) will give \emph{at least\/} the expected error-rate. 
Therefore Alice and Bob might have well shared those particular beneficial
random numbers $r'$ in advance during stage 1 of the protocol. But knowing
$r'$ yields again a deterministic protocol, so the error-rate can also
be reached by a non-randomised (because determined by $r'$) procedure. 
Hence, the bound in Equation (\ref{clas_result}) on deterministic 
protocols translates directly
to non-deterministic protocols as well. 

The second subtlety is on the data distribution side. 
Imagine that the person providing the
numbers $x$ and $y$ is malicious and tries to make the expected error-rate
as high as possible. 
Since he knows Alice and Bob's protocol he can always choose $x$ and 
$y$ so that an error will occur.  Our above result is still safe since 
it is a lower bound, and for the quantum case we already assumed in our
analysis the worst possible distribution on $x$ and $y$. 
However, it is instructive to see that Alice and Bob can take measures 
to keep their error low if they are able to share random numbers secretly. 
There is a large number of different ways of 
colouring the circle for $x$ (i.e. of defining the function $g$).  
We can select those ways that give rise to a minimum error and label them 
with a number $\lambda$.  This number can now serve as a random number 
which is to be shared secretly by Bob and Alice at stage 1. 
In this situation the provider of $x$ and $y$ does not know where 
the boundaries of the black/white colouring are, so he cannot choose 
his numbers $x$ and $y$ to guarantee an error. 
Indeed, his actions will be averaged out and so once again the error 
will be of the order $1/N$. 

In the above described reasonings, two opposing forces are at work. 
i)  If Alice and Bob know the probability distribution over the input 
values, they can optimise their protocol for this case.
ii) For a given probabilistic protocol, the person providing the 
input data can choose the distribution that gives the worst possible
error-rate for the two parties.
Analysing the trade-off between these two forces in a general setting 
finds its origin in game theory and is now also an important part 
of probabilistic computation and communication theory\cite{Kushilevitz}. 

\section{A Multiparty Communication Problem}
\begin{figure}
\begin{picture}(0,0)%
\includegraphics{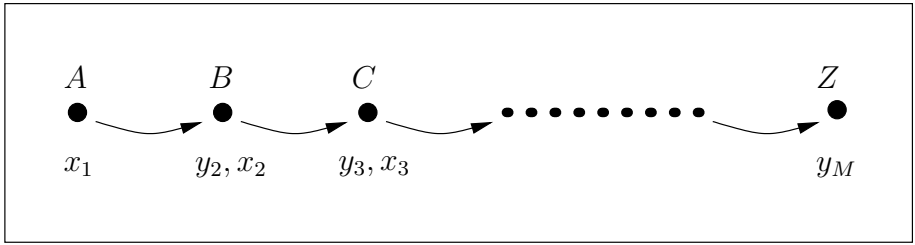}%
\end{picture}%
\setlength{\unitlength}{3947sp}%
\begingroup\makeatletter\ifx\SetFigFont\undefined
\def\x#1#2#3#4#5#6#7\relax{\def\x{#1#2#3#4#5#6}}%
\expandafter\x\fmtname xxxxxx\relax \def\y{splain}%
\ifx\x\y   
\gdef\SetFigFont#1#2#3{%
  \ifnum #1<17\tiny\else \ifnum #1<20\small\else
  \ifnum #1<24\normalsize\else \ifnum #1<29\large\else
  \ifnum #1<34\Large\else \ifnum #1<41\LARGE\else
     \huge\fi\fi\fi\fi\fi\fi
  \csname #3\endcsname}%
\else
\gdef\SetFigFont#1#2#3{\begingroup
  \count@#1\relax \ifnum 25<\count@\count@25\fi
  \def\x{\endgroup\@setsize\SetFigFont{#2pt}}%
  \expandafter\x
    \csname \romannumeral\the\count@ pt\expandafter\endcsname
    \csname @\romannumeral\the\count@ pt\endcsname
  \csname #3\endcsname}%
\fi
\fi\endgroup
\begin{picture}(5724,1524)(-11,-673)
\put(5101,314){\makebox(0,0)[lb]{\smash{\SetFigFont{12}{14.4}{rm}$Z$}}}
\put(376,314){\makebox(0,0)[lb]{\smash{\SetFigFont{12}{14.4}{rm}$A$}}}
\put(2186,314){\makebox(0,0)[lb]{\smash{\SetFigFont{12}{14.4}{rm}$C$}}}
\put(5101,-211){\makebox(0,0)[lb]{\smash{\SetFigFont{12}{14.4}{rm}$y_M$}}}
\put(1286,314){\makebox(0,0)[lb]{\smash{\SetFigFont{12}{14.4}{rm}$B$}}}
\put(376,-211){\makebox(0,0)[lb]{\smash{\SetFigFont{12}{14.4}{rm}$x_1$}}}
\put(1201,-211){\makebox(0,0)[lb]{\smash{\SetFigFont{12}{14.4}{rm}$y_2, x_2$}}}
\put(2101,-203){\makebox(0,0)[lb]{\smash{\SetFigFont{12}{14.4}{rm}$y_3, x_3$}}}
\end{picture}
\caption{Schematic overview of the multiparty whisper protocol: 
a row of parties who are only allowed to communicate like the
two parties in Figure \ref{schematic}.}
\label{multiparty}
\end{figure}

We now come to quantum whispers.  This name is derived from the well 
known game of Chinese whispers in which a row of people pass a message 
from one to next. Typically, this message gets distorted so that the 
message arriving at the end in no way resembles the original 
message.  Thus consider a row of $M$ people $A,B,C,\dots,Z$
(we do not mean to say here that $M=26$). 
Let the number of people be of the order of $N$:
\begin{eqnarray} 
 M & = & cN,
\end{eqnarray} 
where $c$ is a constant (we have in mind a typical value of about $c=10$). 
Now consider the following communication protocol.  In stage 
1 of the protocol the parties are allowed to share information 
pairwise.  Thus, $A$ and $B$ share information, $B$ and $C$ share 
information, and so on.  In the quantum case they are also allowed to 
share entangled states pairwise.  As they did in the two party problem,
they will share singlet states. 
After this $A,B,\ldots,Z$ are given numbers. 
$A$ is given $x_1$ and $B$ is given $y_2$. Also $B$ is given $x_2$
and $C$ is given $y_3$, and so on, so that each party has two numbers 
$y_i$ and $x_i$
except for $A$ (Alice) who has only $x_1$ and $Z$ (Zarah) who has only $y_M$
(see Figure \ref{multiparty}).
All these numbers belong to the set $\{0,1,\ldots,2N-1\}$.
Furthermore the pairs $x_i$ and $y_{i+1}$ satisfy the jump or no jump
condition of Equation (\ref{promise}). (We do not impose any condition on the
pairs $x_i$ and $y_i$.)
In stage 2, $A$ sends one bit ($+1$ or $-1$) to $B$, 
then $B$ sends one bit to $C$, and so on. 
Thus we have the same situation as before but repeated many times
(Figure \ref{multiparty}).
The communication problem is for Zarah, at the end of the row, to 
announce whether there are an even or odd number of jumps. 

\subsection{The Quantum Protocol}
Consider first the quantum case where the parties share singlet states 
pairwise.  
These singlet states consist of a left particle $L$ and a right 
particle $R$. This means Alice has only an $L$ particle, Zarah has only 
an $R$ particle, and all the other parties have both an $R$ and $L$ particle.
Alice measures $2\pi x_1/2N$ on her $L$ particle and Zarah measures 
$2\pi y_M/2N$ on her $R$ particle.
Each of the other parties measures the spin along the direction 
$2\pi y_i/2N$ on their 
$R$ particle and $2\pi x_i/2N$ on their $L$ particle.
As before the sign of the outcome for the measurements on the $R$ particles
is redefined so that $\pm1$ becomes $\mp1$.  
Alice sends the result of her 
measurement to Bob. Bob multiplies this result by both of his results 
and communicates the result of this to $C$ (Carol). 
$C$ multiplies the message she receives from Bob by the results 
of her two spin measurements and communicates this result to $D$. 
This procedure is repeated until Zarah receives a message. 
As a last step she multiplies the message she received from $Y$ by 
the result of her own measurement.  Zarah will now have a number $\pm1$. 
Every jump along the row contributes a factor $-1$, hence if the end result 
is $+1$ then Zarah can announce that there has been an even number of jumps.
If the end result is $-1$ then Zarah can announce that there has been an odd 
number of jumps.  For big $N$ (that is, small $1/N^2$) 
the overall expected error is the sum of the $M-1$ pairwise error
probabilities (which are less than ${(\pi/2N)}^2$).  
Hence we can say that under that assumption,
the overall error is at most $M({\pi/ 2N})^2$ or, using $M=cN$, 
\begin{eqnarray} 
\text{Error}_{\text{quantum whispers}}
& \leq &
\frac{c \pi^2}{4N}~. 
\end{eqnarray} 
Which gives us a useful protocol if $c/N$ is sufficiently small.
 
\subsection{Analysing Possible Classical Protocols}
For the final answer of the protocol, the parties have to solve
$M-1$ independent ``jump or no jump?'' questions. The questions are
independent in the sense that 
Alice's value $x_1$ and Bob's value $y_2$ are uncorrelated with the
other $x$ and $y$ values. Therefore with the one bit of 
communication from Alice to Bob, the two parties have to solve the same 
problem as we discussed in the previous section. Hence Alice
and Bob will have an expected error-rate of the order $1/N$. 
This holds in general for all $M-1$ pairs $(x_i,y_{i+1})$ for
which the jump/no jump question has to be answered.
These errors are uncorrelated with each other.
For small $c$ (meaning $M/N\ll 1$), the overall error-rate is the sum of 
the $M-1$ individual error probabilities, giving
\begin{eqnarray}\label{smallc}
\mathrm{Error}_{\mathrm{classical~whispers}}^{\mathrm{small~}c} & \geq & 
\frac{M-1}{3N}~. 
\end{eqnarray}
For large $c$ this formula is no longer valid.
If we assume (like we did in the quantum case) that $N$ is big,
the expected error-rate will then be:
\begin{eqnarray} 
\mathrm{Error}_{\mathrm{classical~whispers}}^{\mathrm{big~}N} & \geq & 
\frac{1}{2}-\frac{{\mathrm{e}}^{-4c/3}}{2}~. 
\end{eqnarray} 
In the case where $c$ is about 10 this error effectively equals $1/2$.
Thus implying that the protocol is as effective as a random
coin flip by Zarah when choosing her announcement.

The error-rate given for small $c$ (Equation(\ref{smallc})) can actually 
be achieved by the following protocol.
Each pair agrees on a 
function $g$, thus $g_{AB}, g_{BC},\dots$  Alice sends the result 
$g_{AB}(x_1)$ to Bob.  Bob must decide whether he thinks there has been 
a jump or not.  He uses a function 
$h_{AB}(g_{AB}(x_1),y_2)$ 
to decide, where this function has value $+1$ (jump) or $-1$ (no jump). 
The other pairs can define similar functions $h_{BC}$, $h_{CD}$, \dots 
which give the best guess as to whether there has been a jump 
or not.  Now, Bob multiplies $h_{AB}$ by $g_{BC}$ and sends that result 
to Carol.  This proceeds until Zarah receives a message from $Y$ which 
she then multiplies by $h_{YZ}$.  As in the quantum case, Zarah will 
announce that there have been an even number of jumps if the end outcome 
is $+1$ and an odd number of jumps if the end outcome is $-1$.
This protocol becomes unreliable for large $c$ for the reasons given above.

\section{Conclusions}
We have seen how quantum entanglement is useful from a communication 
complexity point of view.  This result employed the very tight quantum 
correlations of a singlet state for small angles. 
First we considered a two party example. The
two parties, Alice and Bob, are given the numbers $x$ and $y$
respectively, where $x,y, \in\{0,\ldots,2N-1\}$.  These two numbers
satisfy either a jump or a no-jump condition 
(defined in Equation (\ref{promise})).
Alice is allowed to communicate one bit of classical information to Bob
and Bob has to announce whether he thinks there has been a jump or not.
In the classical case the error rate goes as $1/N$ whereas in the
quantum case, where the parties are allowed to share an entangled quantum 
state beforehand,  the error goes as $1/N^2$.  Thus as N gets
big the quantum error is significantly smaller.  

We also considered an extension
of this scheme to a situation in which many parties are arranged in a
row.  They are given numbers which, from one party to 
the next in the row, satisfy the jump or no jump condition.  Each party can
communicate one bit of classical information to the next party down the
row.  The object is for the person at the end of the row to announce
whether there has been an odd or an even number of jumps.  We considered
the case where the number of parties in the row is about $10N$. In the
quantum case, in which the parties are allowed to share singlet states
pairwise, the error-rate goes as $1/N$ so is small for large $N$.
However, in the analogous classical scheme the error-rate is
approximately $1/2$.

\section*{Acknowledgements}
LH is funded by the Royal Society.
WvD is supported by the European TMR Research Network 
ERP-4061PL95-1412, Hewlett-Packard, and the Institute for Logic, Language,
and Computation (Amsterdam).


\begin{thebibliography}{99}

\bibitem{Braunstein} 
Samuel L.\,Braunstein and Carlton M.\,Caves,
``Writing out better Bell inequalities''
\emph{Proceedings of the 3rd International Symposium on the Foundations 
of Quantum Mechanics in the Light of New Technology,\/} 
editors S.\,Kobayashi \emph{et al\/}, Physical Society of Japan,
pp.~161--170 (1990).

\bibitem{Buhrman1} 
Harry Buhrman, Richard Cleve, and Wim van Dam,
``Quantum entanglement and communication complexity'', 
preprint on the quant-ph archive\footnote{
The quant-ph archive can be found at: 
\texttt{http://xxx.lanl.gov/archive/quant-ph/}}, no.~9705033 (1997).

\bibitem{Buhrman2} Harry Buhrman, Richard Cleve, and Avi Wigderson,
``Quantum vs. Classical Communication and Computation'',
to appear in 
\emph{Proceedings of the 30th Annual ACM Symposium on Theory of 
Computing,\/} ACM Press (1997). 
Also as preprint on the quant-ph archive, no.~9705033.

\bibitem{Cleve1} 
Richard Cleve and Harry Buhrman,
``Substituting quantum entanglement for communication'', 
\emph{Physical Review A\/} \textbf{56(2)}, pp.~1201--1204 (1997). 
Also as preprint on the quant-ph archive, no.~9704026.

\bibitem{Cleve2} 
Richard Cleve, Wim van Dam, Michael Nielsen, and Alain Tapp,
``Quantum Entanglement and the Communication Complexity of the Inner Product 
Function'', 
to appear in 
\emph{Proceedings of the 1st NASA International Conference on Quantum 
Computing and Quantum Communications,\/} Springer-Verlag (1998),
Also as preprint on the quant-ph archive, no.~9708019.

\bibitem{Cover}
Thomas M.\,Cover and Joy A.\,Thomas,
\emph{Elements of Information Theory,\/}
John Wiley \& Sons, Inc. (1991).

\bibitem{vDam} 
Wim van Dam, Peter H{\o}yer, and Alain Tapp,
``Multiparty Quantum Communication Complexity'',
preprint on the quant-ph archive, no.~9710054 (1997).

\bibitem{Ekert}
Artur Ekert, Susana Huelga, Chiara Macchiavello, Juan Ignacio Cirac,
``Distributed Quantum Computation over Noisy Channels'',
preprint on the quant-ph archive, no~9803017 (1998).

\bibitem{Grover} 
Lov K. Grover,
``Quantum telecomputation'',
preprint on the quant-ph archive, no.~9704012 (1997).

\bibitem{Hardy} 
Lucien Hardy, 
``A new way to obtain Bell inequalities''
\emph{Physics Letters A\/} \textbf{161}, pp.~21--25 (1991).

\bibitem{Kushilevitz}
Eyal Kushilevitz and Noam Nisan,
\emph{Communication Complexity,\/} Cambridge University Press (1997).

\bibitem{Squires} 
Euan J.\,Squires, Lucien Hardy, and Harvey R.\,Brown,
``Nonlocality from an analogue of the quantum Zeno effect'',
\emph{Studies in History and Philosophy of Science\/} \textbf{25}, 
pp.~425--435 (1994).

\bibitem{Werner}
Reinhard F.\,Werner,
\emph{Quantentheorie,\/}
unpublished lecture notes in German, available at:
ftp://ftp.physik.uni-osnabrueck.de/pub/werner/

\end{thebibliography}
\end{document}